\documentclass[aps,epjc,twocolumn,floatfix,superscriptaddress,nofootinbib,showpacs,showkeys,preprintnumbers]{revtex4}
\usepackage{graphicx}
\usepackage{dcolumn}  
\usepackage{bm}       
\usepackage{palatino}
\usepackage{color}

\begin{document}
\input{epsf}
\Large
\title{ Possible influence of the two string events on the
    hadron formation in a nuclear environment}

\normalsize

\author{N.~Akopov}
\author{L.~Grigoryan}
\author{Z.~Akopov,\\
Yerevan Physics Institute, Br.Alikhanian 2, 375036 Yerevan, Armenia}

\begin {abstract}
\hspace*{1em} 
One of the basic assumptions of the string model is that as a result of 
a DIS in nucleus a single string arises, which then breaks into
hadrons. However the pomeron exchange considered in this work, leads to the
production of two strings in the one event.
The hadrons produced in these events have smaller formation lengths, than those with the same 
energy produced in the single string events. As a consequence, they 
undergo more substantial absorption in the nuclear matter.
\end {abstract}
\maketitle
\section{Introduction}
\normalsize
\hspace*{1em} Production of hadrons in a nuclear environment is a well
known tool for investigations of the early stage of hadronization.
The basic assumption of the parton model is that the virtual photon 
interacts as a point-like electromagnetic probe. 
Being directly connected with the charge of the single parton in the target,
the virtual photon knocks it out, transferring it's energy. 
A color string that is stretched between the knocked out parton and the target
remnant then breaks into the final hadrons.
(see Fig.1).
\begin{figure}[!hbt]
\begin{center}
\epsfxsize=4.cm
\epsfbox{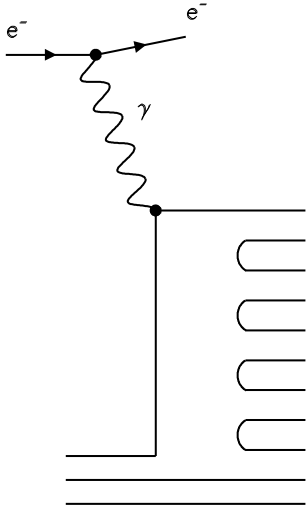}
\end{center}
\caption{\label{xx1}
{\it
The single string mechanism of the hadrons electroproduction.
The virtual photon interacts as point-like electromagnetic probe
with the parton in target, knocks it out. 
 }}
\end{figure}
In this article we consider an additional mechanism connected with the possibility of a 
simultaneous production of two strings.
This happens due to the complicated nature of the photon\footnote{
Photon has properties of an electromagnetic
probe and hadron and interacts with a hadrons via three types of processes:
direct, VMD, and anomalous. }
, which can interact as a hadron-like system consisting of a quark and
an antiquark with a gluonic field of the nucleon (pomeron) and
by means of quark-antiquark exchange (reggeons).
Diagrams describing the mechanism of the high energy hadron-hadron
interaction in Regge theory are shown in Fig.2. The surfaces of planar and cylindrical diagrams are covered by gluons
and quark-antiquark pairs, which do not represented for the sake of
simplicity.
\begin{figure}[!hbt]
\begin{center}
\epsfxsize=6.cm 
\epsfbox{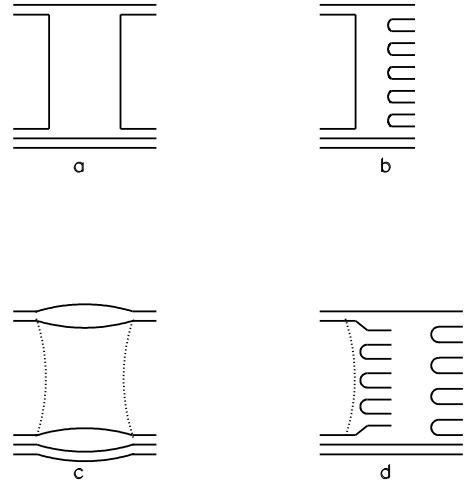}
\end{center}
\caption{\label{xx1}
{\it
The hadron-hadron
interaction in the Regge theory. The interaction
via exchange of reggeon (Fig. 2a) and pomeron (Fig. 2c) in
elastic scattering is shown. 
Contributions of reggeons and 
pomeron in a multi-particle   
production are shown in Figs.2b and 2d. 
 }}
\end{figure}

One should note, that the reggeon exchange leads
to a one string picture as in Fig.1, whereas pomeron exchange leads
to a two string mechanism of hadron production.
The main goal of this work is to study the contribution of the two string
events in an electro-production on nuclei. 
For this purpose we investigate the nuclear attenuation (NA), which
is the ratio of the differential multiplicity on a nucleus (A) to that on
deuterium (D).
The experimentally measured observable, NA, is usually considered as a function 
of three kinematical variables:
energy of the virtual photon $\nu$, fraction of photon's energy carried by
final hadron $z=E_{h}/\nu$, where $E_{h}$ is the energy of hadron in laboratory
system and square of the four momentum of photon $q^{2} = - Q^{2}$
\begin{eqnarray}
\nonumber
R^{h}_{M}(\nu, z, Q^{2}) = \frac{\Big(\frac{N^{h}(\nu,z,Q^{2})}{N^{e}(\nu,Q^{2})}
\Big)_{A}}{\Big(\frac{N^{h}(\nu,z,Q^{2})}{N^{e}(\nu,Q^{2})}\Big)_{D}} ,
\end{eqnarray}
with $N^{h}(\nu,z,Q^{2})$ being the number of semi-inclusive hadrons in a given
$(\nu,z,Q^{2})$ bin, and $N^{e}(\nu,Q^{2})$ being the number of inclusive
DIS leptons in the same $(\nu,Q^{2})$ bin. However, we consider
$R^{h}_{M}$ as a function of two variables, $(\nu,Q^{2})$ or $(z,Q^{2})$,
which assumes that integration over the third kinematic variable is done.
As it has been shown, the contribution of the two string mechanism in $R^{h}_{M}$
is negligible at $Q^2 > 10GeV^2$.
It is in order of few percent of the basic (single string)
mechanism in case of the HERMES kinematics ($Q^2 \approx 2.5 GeV^2$).
However it can increase essentially at lower values of $Q^2$.\\
The paper is organized as follows. In Section 2 we briefly discuss how
the fraction of two string events can be obtained. 
Formulae for the calculation of the virtual photon energy division between two
strings are presented in Section 3.
Section 4 presents the results and discussion. Our conclusions are
presented in Section 5.  
\section{Fraction of two string events}
\normalsize
\hspace*{1em}  
Total hadronic cross sections show a characteristic fall-off at low
energies and a slow rise at higher energies. This behavior can be
parametrized by the form~\cite{A1}
\begin{eqnarray}
\sigma^{AB}_{tot}(s)=X^{AB}s^{\epsilon} + Y^{AB}s^{-\eta}
\hspace{0.2cm}[mb]
\end{eqnarray}
for $A + B \to X$ and $s = (p_A + p_B)^2$, with s in $GeV^2$.
 The powers $\epsilon$ and $\eta$ are universal, with
 fit values
\begin{eqnarray}
\epsilon \approx 0.0808,  \hspace{0.3cm}\eta \approx 0.4525,
\end{eqnarray}  
while the coefficients $X^{AB}$ and $Y^{AB}$ are process-dependent.
Equation (1) can be interpreted within Regge theory, where the first term 
corresponds to the pomeron exchange and gives the asymptotic rise of the cross
section. The second term, the reggeon one, is mainly of interest at low
energies. Empirically, the $\gamma p$ data are well described by the
s-dependence of type eq. (1)
\begin{eqnarray}
\sigma^{\gamma p}_{tot}(s) \approx 67.7s^{\epsilon} + 129s^{-\eta}
\hspace{0.2cm}[\mu b],
\end{eqnarray}
Actually, the above formula is a prediction~\cite{A1}
preceding the HERA data~\cite{A2}. 
On the other side,
it is well known that the total $\gamma p$ cross section may be written as
consisting from three parts (see for instance~\cite{A3})
\begin{eqnarray}
\sigma^{\gamma p}_{tot} = \sigma^{\gamma p}_{VMD} + \sigma^{\gamma
p}_{dir} + \sigma^{\gamma p}_{anom},
\end{eqnarray}
where the subdivision of total cross section corresponds to 
 the existence of three main event classes in
$\gamma p$ events:\\
1. The Vector Meson Dominance (VMD) processes, 
where the photon turns into a vector meson before
 the interaction with target, and therefore all processes allowed in
hadronic physics may occur.\\
2. The direct processes, where a bare photon interacts with a parton 
from the proton.\\
3. The anomalous processes, where the photon perturbatively branches
into a $q\overline{q}$ pair, and one of these interacts with a parton
from the proton.\\
The total VMD cross section is obtained as weighted sums of the
allowed vector-meson states~\cite{A3,A4}
\begin{eqnarray}
\sigma^{\gamma p}_{VMD}=\sum_{V=\hspace{0.1cm}\rho^{0},
\hspace{0.1cm}\omega, \hspace{0.1cm}\phi}
\frac{4\pi\alpha_{em}}{f^{2}_{V}}
\sigma^{Vp}_{tot},
\end{eqnarray}
where $\alpha_{em}  \approx 1/137$,\hspace{0.2cm}$f^{2}_{V}/4\pi$
determined from 
data to be 2.20 for $\rho^{0}$, 23.6 for $\omega$ and 18.4 for
$\phi$~\cite{A4}, $\sigma^{Vp}_{tot}$ are corresponding vector
meson-proton total cross sections, which can be find, for instance
in Ref.~\cite{A3}.\\
Now, let us turn to the electro-production process, when one is dealing 
with
the virtual photon interaction. The phenomenological model which extends
VMD to the case of the virtual photon is the Generalized VMD (GVMD)~\cite{A5}. 
In this case we have a more complicated connection with the vector meson
total cross sections
\begin{eqnarray}
\sigma^{\gamma^{*} p}_{GVMD}=\sum_{V=\hspace{0.1cm}\rho^{0}, 
\hspace{0.1cm}\omega, \hspace{0.1cm}\phi}
\frac{4\pi\alpha_{em}}{f^{2}_{V}}\frac{1}{(1 + Q^{2}/m^{2}_{V})^{2}}
\sigma^{Vp}_{tot},
\end{eqnarray}
where $\sigma^{\gamma^{*} p}_{GVMD}$ is function of two variables:
the c.m. energy of the $\gamma^{*} p$ system $W$ and $Q^{2}$.
\hspace*{1em}
The total $\gamma^{*} p$ cross section, $\sigma^{\gamma^{*} p}_{tot}$,
can be related to the proton structure function $F_{2}$ through
the relation (see, for instance, Ref.~\cite{A6} )
\begin{eqnarray}
\sigma^{\gamma^{*} p}_{tot}(x,Q^{2}) = \frac{4\pi^{2}\alpha_{em}}
{Q^{2}(1 - x)}\frac{Q^{2} + 4m^{2}_px^{2}}{Q^{2}}F_{2}(x,Q^{2}),
\end{eqnarray}
where the total $\gamma^{*} p$ cross section includes both the cross
section for the absorption of transverse and of longitudinal photons,
$x$ is Bjorken variable $x=Q^{2}/2m_{p}\nu$, where $m_{p}$ is 
the proton mass. 
Proton structure function $F_{2}$ was calculated using a model presented
in Ref.~\cite{A7}.
Using parameterization (1) for $\sigma^{Vp}_{tot}$ for $V = \rho^{0}, 
\omega, \phi$ as presented in Ref.~\cite{A3},
the formula (6) can be written separately for the contribution of pomeron in
$\gamma^{*} p$ total cross section $\sigma^{\gamma^{*} p(P)}_{GVMD}$.
The knowledge of this quantity allows one to obtain the relative share of 
the contribution of the pomeron in the virtual photon - proton total cross section
\begin{eqnarray}
\alpha = \frac{\sigma^{\gamma^{*} p(P)}_{GVMD}}{\sigma^{\gamma^{*}
p}_{tot}}.
\end{eqnarray}
\hspace*{1em}The Regge theory is widely used to describe the low $p_{T}$
high-energy interactions of hadrons, nuclei, real and virtual photons.
The Pomeranchuk singularity plays a special role in this theory as it determines
the high energy behaviour of diffractive processes and multi-particle
production. It is possible to connect the general results of the Regge
theory with the parton model.
We will follow the representation of Dual Parton Model
(DPM)~\cite{A8}, which was developed by incorporating partonic ideas
into dual topological unitarization\footnote{Similar ideas lie in the
base of Quark Gluon String Model~\cite{A9}}. 
In DPM a meson (baryon) are a excitations of an open
string with valence quark and antiquark (diquark) at its ends. When
a string is stretched, it decays into hadrons by breaking into short
strings. The dominant contribution to the high energy scattering of
two hadrons comes from a closed string (a pomeron) exchange, having a
cylinder topology (see Fig.2c). A unitarity cut of the cylindrical pomeron
shows
that the sources of the multi-particle production are two hadronic chains
(see Fig.2d).
Other reggeons correspond to the planar diagrams, which give one
hadronic chain in the multiparticle production.
The presence of contribution from pomeron leads to the arising
of the two string events in DIS in addition to the single string events.
The relative share of these events given by eq. (8).

\section{Energy division between two strings}
\normalsize
\hspace*{1em}
In the two string events the energy of the projectile (virtual photon) 
is shared by two strings. We will use two methods for the definition of
this energy division.\\
As a first method, we will use the functions $W_{T,L}(\beta, r_{T})$
from Ref.~\cite{A10}, which play a role of the square of the wave functions
of the virtual photon's quark-antiquark fluctuations. In an explicit form:
\begin{eqnarray}
W_{T}(\beta, r_{T}) = \frac{6\alpha_{em}}{(2\pi)^{2}}\sum^{N_f}_{i=1}
e^{2}_{i}\{[1-2\beta(1-\beta)]\varepsilon^{2}K^{2}_{1}(\varepsilon r_{T}) +
\nonumber
\end{eqnarray}
\begin{eqnarray}
 m^{2}_{i}K^{2}_{0}(\varepsilon r_{T}) \}
\end{eqnarray}
and
\begin{eqnarray}
W_{L}(\beta, r_{T}) = \frac{6\alpha_{em}}{(2\pi)^{2}}\sum^{N_f}_{i=1} 
e^{2}_{i}4Q^{2}\beta^{2}(1-\beta)^{2}K^{2}_{0}(\varepsilon r_{T}),
\end{eqnarray}
where $m_{i}$ is the mass of the quark $i$, $e_{i}$ is the quark charge,
$\beta$ is the fraction of the $q\bar{q}$ momentum carried by one of the
quarks, $\varepsilon^{2} = m^{2}_{i} + \beta(1-\beta)Q^{2}$, $r_{T}$ is 
the transverse size of the $q\bar{q}$ pair, $K_{0,1}(x)$ are modified
Bessel functions. It is impossible to normalize the virtual photon wave function 
to unity because the normalization integral
\begin{eqnarray}
N_{\gamma^{\ast}} = \int^{1}_{0}d\beta\int d^{2}r_{T}W_{T}(\beta, r_{T})
\end{eqnarray}
diverges logarithmically at small distances, since $K_{1}(x) \sim 1/x$ at
$x \rightarrow 0$. This divergence does not cause any problems if one includes
in the normalization integral the dipole-nucleon cross section, since 
$\sigma_{q\bar{q}}(r_{T}) \sim r^{2}_{T}$ at small $r_{T}$. Then one obtains the
probability function for the energy division between two strings normalized to
unity in form:
\begin{eqnarray}
\nonumber
{w(\beta) = \int d^{2}r_{T}\sigma_{q\bar{q}}(r_{T})[W_{T}(\beta, r_{T}) +
\epsilon W_{L}(\beta, r_{T})]/    \hspace{2.0cm}}
\end{eqnarray}
\begin{eqnarray}
{\int^{1}_{0}d\beta\int d^{2}r_{T}\sigma_{q\bar{q}}(r_{T})
[W_{T}(\beta, r_{T}) + \epsilon W_{L}(\beta, r_{T})]},
\end{eqnarray}
where $\epsilon$ is photon polarization and $\sigma_{q\bar{q}}(r_{T})$, as
mentioned above, is the cross section for the dipole-nucleon interaction.
\footnote{For calculations we use $\sigma_{q\bar{q}}(r_{T})$ in the form presented
in Ref.~\cite{A11} by formulae (9)-(12). The only difference is that for the 
pion-nucleon total cross section we use a constant value $\sigma^{\pi p}_{tot} =25mb$.} 
\\ 
As the second method we will use a probability function for the energy division 
$w(y_{q}, y_{\bar{q}})$ in form presented in Ref.~\cite{A12}. This function can be
applied to any hadronic system that consists of a quark and antiquark
\begin{eqnarray}
w(y_{q}, y_{\bar{q}}) = Cexp(-(1 - \alpha_R(0))|y_{q} - y_{\bar{q}}|),
\end{eqnarray}
where $y_{q}$ and $y_{\bar{q}}$ are rapidities of the quark and the antiquark,
respectively, $\alpha_R(0)$ is an intercept of the secondary Regge pole
 $\alpha_R (R = \rho, \omega, f)$. The commonly used average value for $\alpha_R 
(0)$ is $\alpha_R(0) \approx 0.5$, C is a normalization factor.
For calculations it is convenient to present the function $w(y_{q}, y_{\bar{q}})$ 
in form:
\begin{eqnarray}
w(\beta) \approx C\Bigl[ \frac{min(\beta,1-\beta)}{max(\beta,1-\beta)}\Bigr]^{(1 -
\alpha_R(0))}.   
\end{eqnarray}
Strictly speaking, the energy division function in form eq.(13)-(14) is suitable
for the region $|y_{q} - y_{\bar{q}}| \gg 1$, but in this paper, 
taking into account its qualitative character, we will use this formula
for the full region of $y_{q} - y_{\bar{q}}$. We will discuss such choice
later in Results and Discussion.

\section{Results and Discussion}
\normalsize
\hspace*{1em}
Using quantities defined in the preceding sections and the formalism of 
Ref.~\cite{A13}, we can now calculate the NA,
taking into account the admixture of the two string events.
In order to do this, we must change $R_A$ in eq.(1) Ref.~\cite{A13}, to 
$(1 - \alpha ) R_A + 2 \alpha R^{'}_{A}$, 
where $R_{A}$ and $R^{'}_{A}$ are the absorption functions for the
hadron produced in the single string event and in
one of the strings of the two string event, respectively.
$R^{'}_{A}$ is integrated over $\beta$ with energy division functions
$w(\beta)$ presented by equations (12) and (14).
\begin{figure}[!hbt]
\begin{center}
\epsfxsize=8.cm
\epsfbox{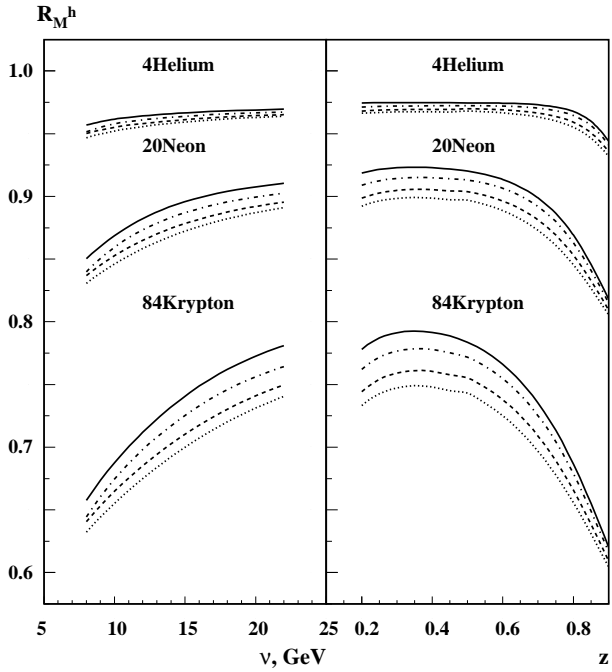}
\end{center}
\caption{\label{xx1}
{\it NA ratio for pions on different nuclei as a
functions of $\nu$ (left panel) and $z$ (right panel) in the
framework of TSM with $\tau_c$ from Lund model.
Solid curves represent single string case;
other curves represent cases with admixture of two string events.
Fraction of two string events depends from the value of $Q^2$.
Dotted curves correspond $Q^2=1GeV^2$, dashed $Q^2=2.5GeV^2$ and
dashed-dotted $Q^2=10GeV^2$.
In calculation was used function $w(\beta)$ from eq.(12).
 }}
\end{figure}

\begin{figure}
\begin{center}
\epsfxsize=8.cm
\epsfbox{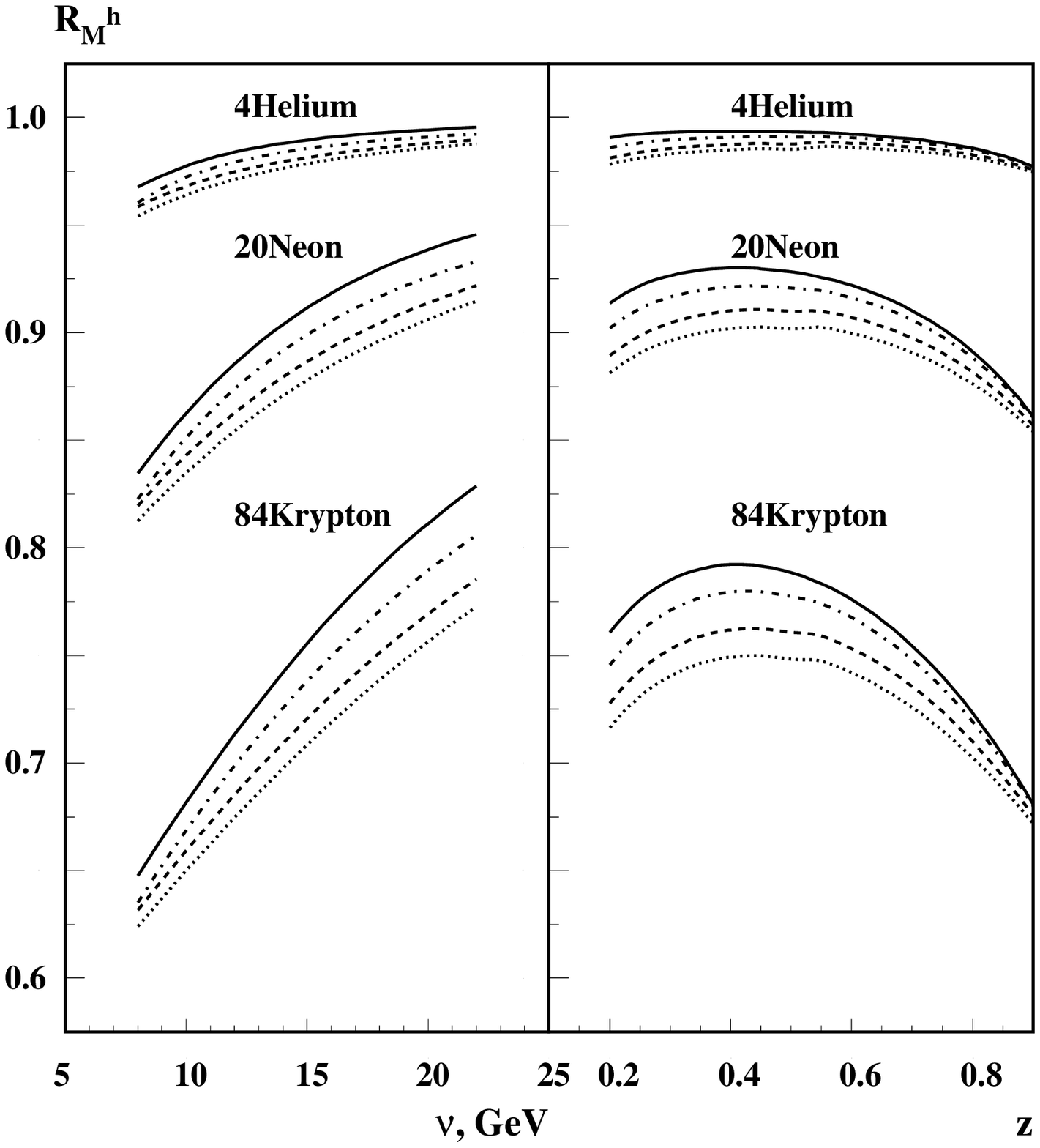}
\end{center}
\caption{\label{xx2}
{\it NA ratio for pions on different nuclei as a
functions of $\nu$ (left panel) and $z$ (right panel) in the
framework of ITSM with $\tau_c$ from Lund model.
The rest as in caption of Fig.3
 }}
\end{figure}

\begin{figure}[!hbt]
\begin{center}
\epsfxsize=8.cm
\epsfbox{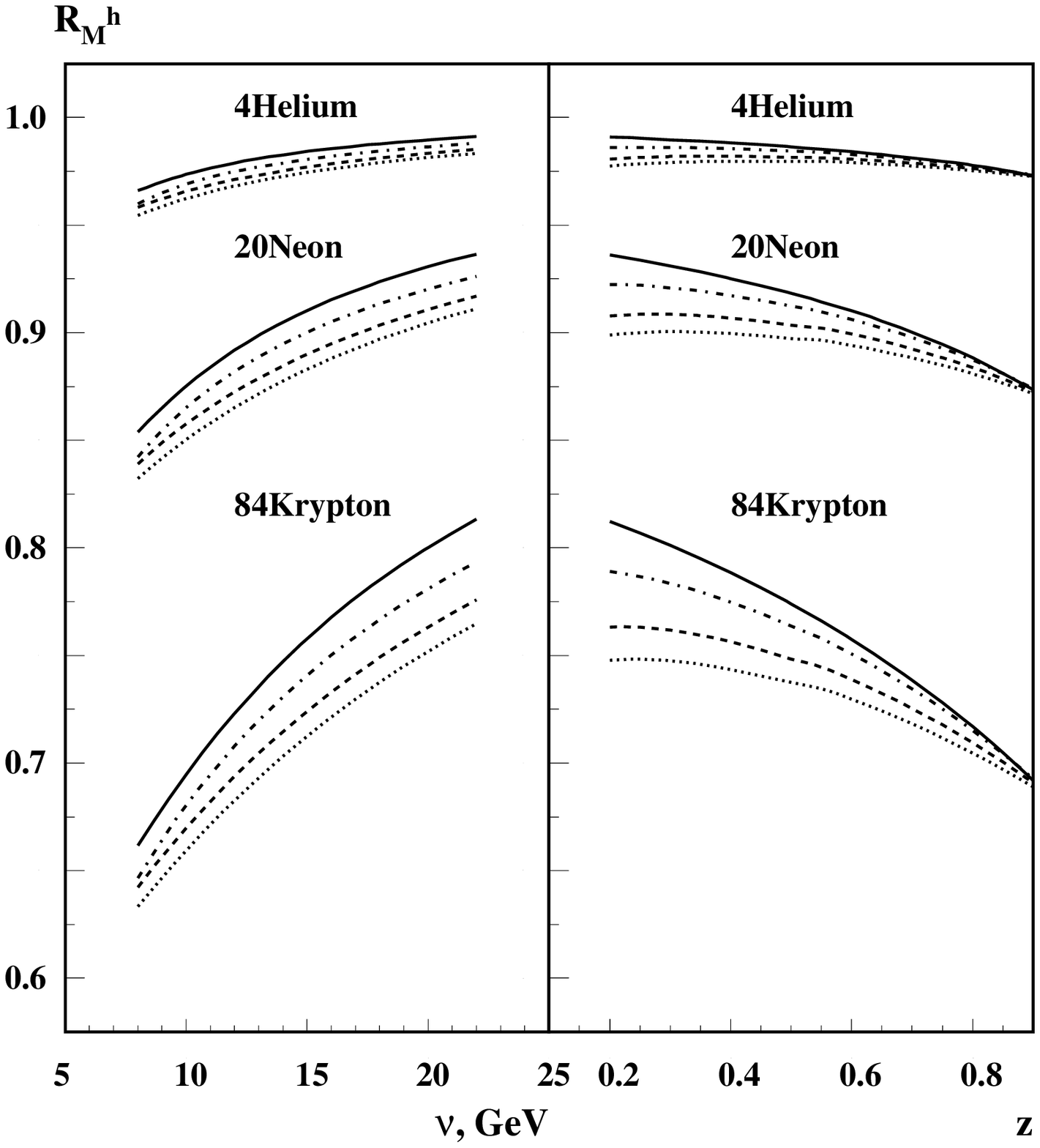}
\end{center}
\caption{\label{xx1}
{\it NA ratio for pions on different nuclei as a
functions of $\nu$ (left panel) and $z$ (right panel) in the
framework of ITSM with $\tau_c$ for leading hadron (see [13]).
The rest as in caption of Fig.3
 }}
\end{figure}

\begin{figure}[!hbt]
\begin{center}
\epsfxsize=8.cm
\epsfbox{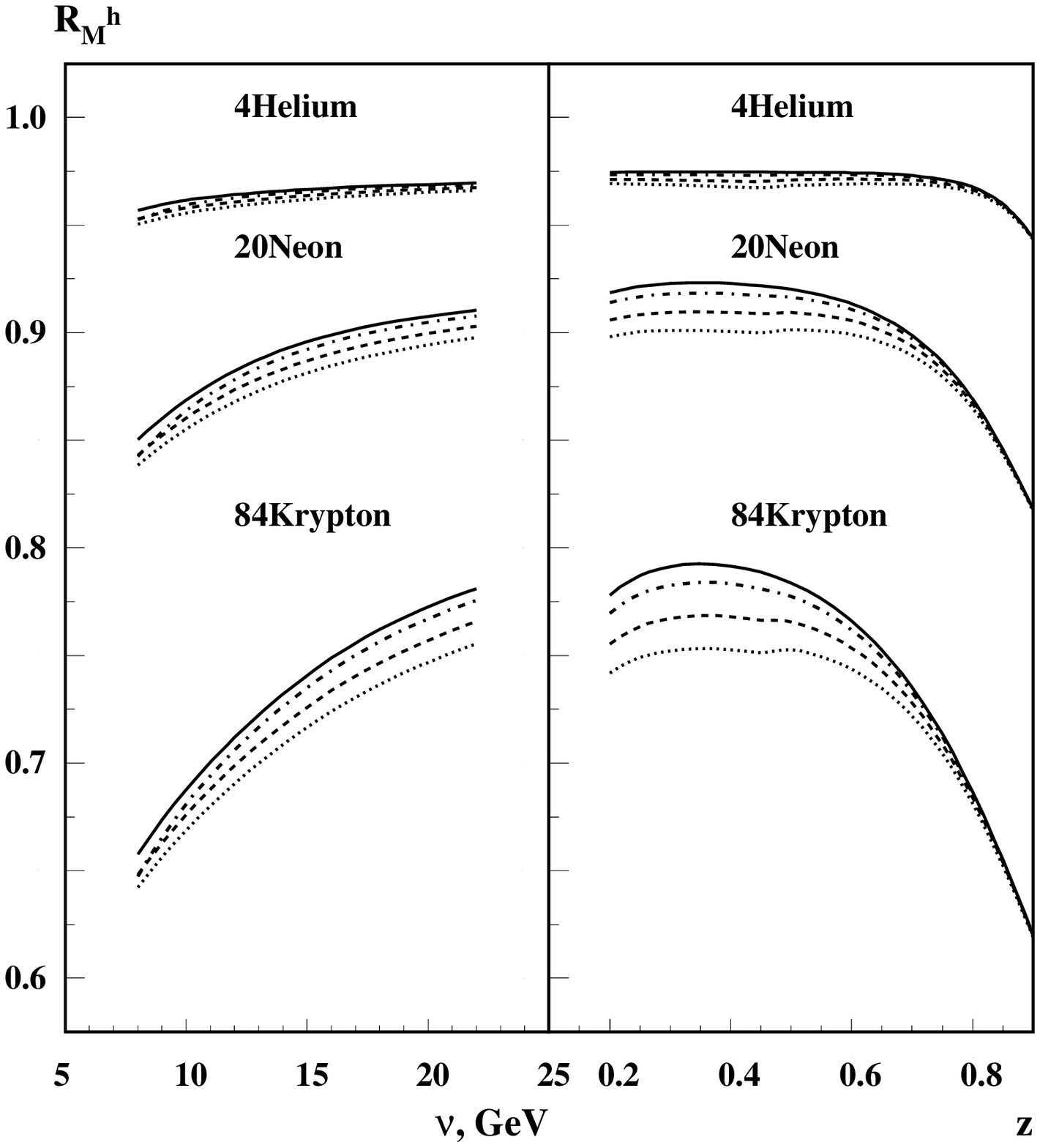}
\end{center}
\caption{\label{xx1}
{\it NA ratio for pions on different nuclei as a
functions of $\nu$ (left panel) and $z$ (right panel) in the
framework of TSM with $\tau_c$ from Lund model.
Solid curves represent single string case;
other curves represent cases with admixture of two string events.
Fraction of two string events depends from the value of $Q^2$.   
Dotted curves correspond $Q^2=1GeV^2$, dashed $Q^2=2.5GeV^2$ and 
dashed-dotted $Q^2=10GeV^2$.
In calculation was used function $w(\beta)$ from eq.(14).
 }}
\end{figure}

\begin{figure}[!hbt]
\begin{center}
\epsfxsize=8.cm
\epsfbox{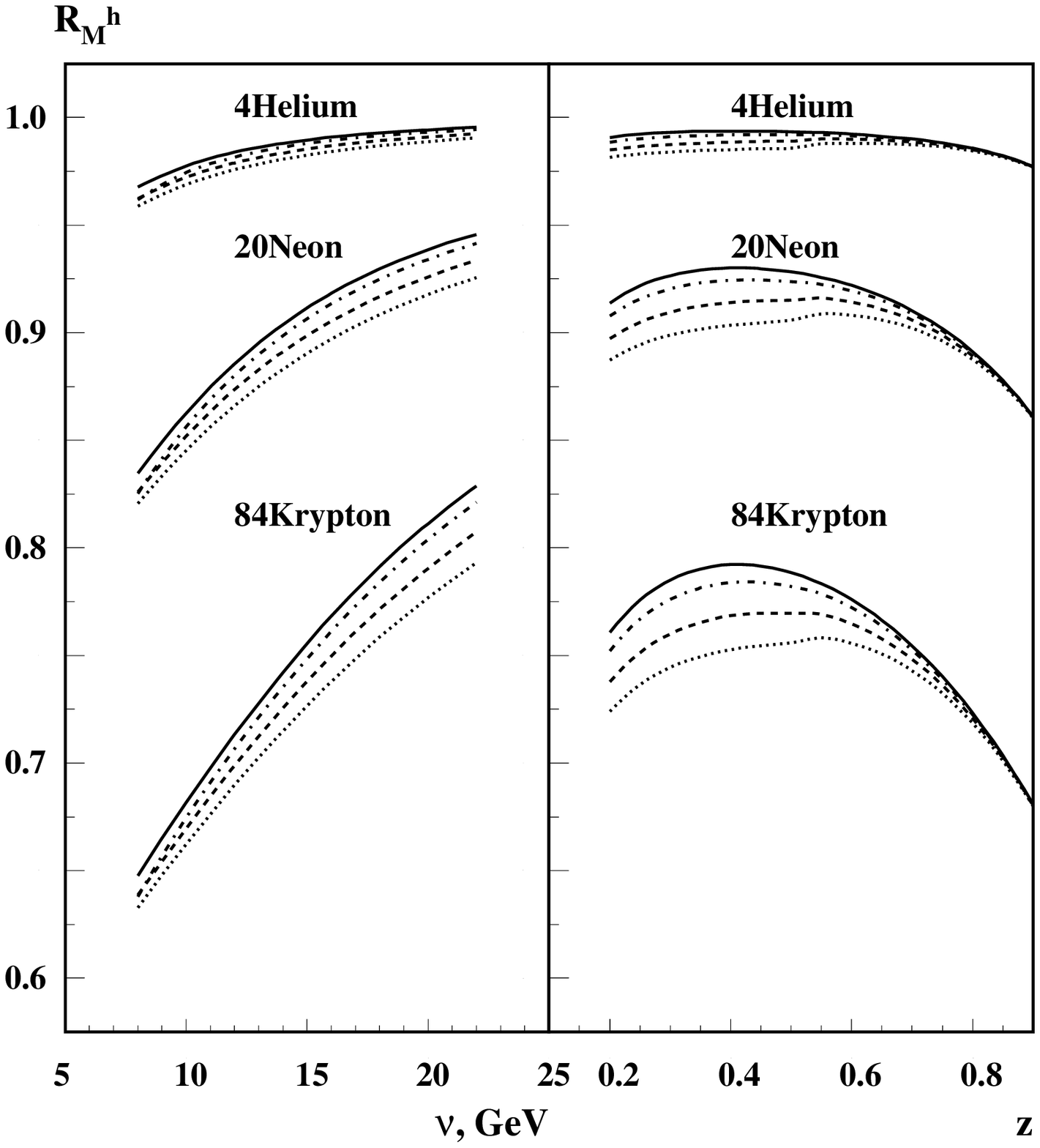}
\end{center}
\caption{\label{xx2}
{\it NA ratio for pions on different nuclei as a
functions of $\nu$ (left panel) and $z$ (right panel) in the
framework of ITSM with $\tau_c$ from Lund model.
The rest as in caption of Fig.6.
 }}
\end{figure}
\begin{figure}[!t]
\begin{center}
\epsfxsize=8.cm
\epsfbox{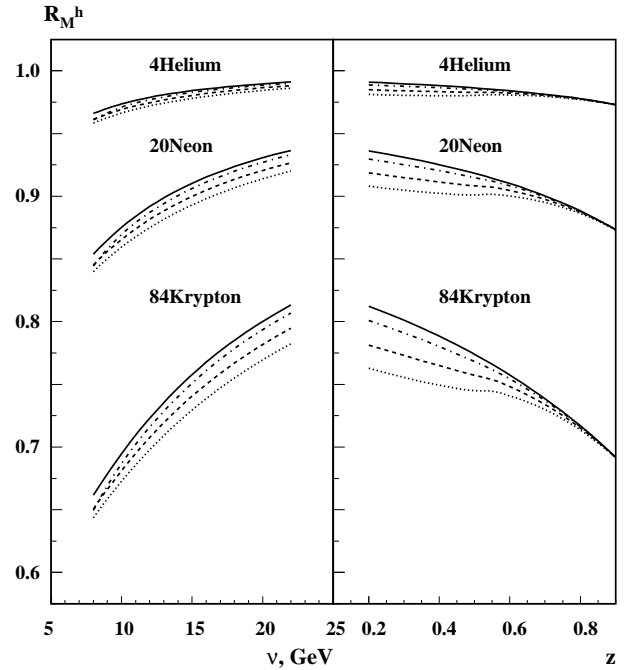}
\end{center}
\caption{\label{xx1}
{\it Nuclear attenuation ratio for pions on different nuclei as a 
functions of $\nu$ (left panel) and $z$ (right panel) in the
framework of ITSM with $\tau_c$ for leading hadron (see [13]).
The rest as in caption of Fig.6.
 }}
\end{figure}
Now, following the formalism of~\cite{A13} we can write the NA in form:
\begin{eqnarray}
R^{h}_{M}(\nu, z, Q^{2}) = \frac{(1 - \alpha ) R_A + 2 \alpha R^{'}_{A}}
{(1 - \alpha ) R_D + 2 \alpha R^{'}_{D}},
\end{eqnarray} 
where $R_{D}$ and $R^{'}_{D}$ are the absorption functions for deuterium.
The NA for pions on helium, neon and krypton nuclei as a
functions of $\nu$  and $z$ calculated with $w(\beta)$
obtained from virtual photon wave functions (see eq.(12)), 
are presented in Figs.3 - 5.
In Figs.6 - 8 the same results are obtained with $w(\beta)$ in form eq.(14).
On Figs.3 and 6 the NA in the
framework of TSM with $\tau_c(4)$ (Lund model)\footnote{ 
In this paper the models and the constituent formation times
$\tau_c$ are marked as in Ref.~\cite{A13}. In particular, $\tau_c(3)$
and $\tau_c(4)$ means that $\tau_c$ is taken in form eqs.(3) and (4)
of the referred paper.} are used.
Solid curves represent single string case, while other curves represent 
cases with admixture of the two string events.
The fraction of the two string events depends from the value of $Q^2$.
Dotted curves correspond $Q^2=1GeV^2$, dashed $Q^2=2.5GeV^2$ and
dashed-dotted $Q^2=10GeV^2$.
In Figs. 4 and 7 the NA in the framework of the ITSM with $\tau_c(4)$ are presented.
Figures 5 and 8 show the NA in the
framework of ITSM with $\tau_c(3)$ (leading hadron).
Let us now discuss the results of calculations. First of all
one should note several common features of the results:
a) admixture of two string events always increases the effect of
nuclear attenuation, i.e. increases gap between unity and value of
$R^{h}_{M}$;
b) contribution of the two string events increases
with the increase of the atomic mass number;
c) contribution of the two string events also increases             
with larger values of $\nu$, and decreases 
with the larger values of $z$;
d) $z$-dependence is more sensitive to the contribution of the two 
string events than the $\nu$-dependence.
From Figs. 3 to 8 one can see that the contribution of the two string events is
negligible for all nuclei at $Q^2 > 10GeV^2$. At moderate $Q^2$
it becomes more essential. For instance at $Q^2 = 2.5GeV^2$,
which corresponds to the HERMES kinematics, the contribution of the two string 
events in $z$-dependence on Krypton achieves ~ 3-4\%. For $Q^2 = 1GeV^2$
it increases to 5-6\% (As an observable we use the NA ratio $R^{h}_{M}$.
If nuclear effects are absent, it is obviously equal to unity. The quantity directly
connected with the nuclear effects is $1 - R^{h}_{M}$. The relative contribution 
of the two string events in this quantity is several times larger than
in $R^{h}_{M}$.).
Fig. 3 to 8 also show that the contribution of the two string events is
always small for helium.
Comparing Figs. 3 - 5 with Figs. 6 - 8 one can see that the calculations with functions
$w(\beta)$ defined in equations (12) and (14) give results that are close in shape,
however the contribution of two string events calculated with $w(\beta)$ 
as in eq.(12) is more significant numerically.
We have performed an additional test for the probability function of the energy division $w(y_{q}, 
y_{\bar{q}})$. Besides the eq.(14) we also used an expression $w(y_{q}, y_{\bar{q}})$ =
const; and also a combined expression when constant value of $w$ was used for close values of
$y_{q}$ and $y_{\bar{q}}$ and eq.(14) was used for the case  
$|y_{q} - y_{\bar{q}}| \gg 1$. Changing the expressions 
for $w$ leads to slight change of the form of the curves, but
does not change the contribution level significantly.
In this paper we have limited ourself to consideration of NA for pions
only, but the described mechanism of the two string admixture is common
for all mesons and baryons, and extension is quite straightforward.
We have studied the two string admixture in the framework of the string model,
but in any model that deals with the NA there is a need to take into account the fact that
the virtual photon can transfer its energy both single parton and to two
partons.
The last point which we would like to discuss is what happens if, 
instead of a single string fit of the NA data (performed in~\cite{A13}),
we perform a fit including two string admixture to the single string
mechanism. We do not expect that the numerical values of the fitting parameters
and $\chi^2$ will change significantly. Fitting parameters $\sigma_q$
and $\sigma_s$ in TSM and $\sigma_q$ and c in ITSM, should
become slightly smaller (in order of $\sim 10\%$), and all curves should 
have a somewhat smoother behavior (definitions of free parameters see in Ref.~\cite{A13}).

\section{Conclusions}
\begin{itemize}
\item {
For the first time the contribution of 
two string events in the NA was investigated.
Admixture of the two string events always increases the effect of
NA, i.e. increases gap between unity and $R^{h}_{M}$ value.
}
\item {
Contribution of two string events increases
with the increasing of the atomic mass number; as well as
with the increasing of $\nu$, and decreases
with the increasing of $z$.
It is more prominent in the $z$-dependence.
}
\item {
At moderate $Q^2$ contribution of the two string events is essential
enough.
For instance in the $z$-dependence on krypton it is approximately 5-6\%
at $Q^2 = 1GeV^2$,  ~ 3-4\% at $Q^2 = 2.5GeV^2$. 
It is negligible for all nuclei at $Q^2 > 10GeV^2$. 
}
\item {
It is very interesting to investigate the contribution of
two string events in case of double hadron production~\cite{A14},~\cite{A15}.
}
\end{itemize}


\end{document}